# Adaptive optical focusing through perturbed scattering media with dynamic mutation algorithm

HUANHAO LI[1,2†], CHI MAN WOO[1,2†], TIANTING ZHONG[1,2], ZHIPENG YU[1,2], YUNQI LUO[3], YUANJIN ZHENG[3], XIN YANG[4], HUI HUI[4*], AND PUXIANG LAI[1,2*]

[1]*Department of Biomedical Engineering, Hong Kong Polytechnic University, Hong Kong SAR, China*
[2]*Hong Kong Polytechnic University Shenzhen Research Institute, Shenzhen, Guangdong, China*
[3]*School of Electrical and Electronics Engineering, Nanyang Technological University, Singapore*
[4]*CAS Key Laboratory of Molecular Imaging, Institute of Automation, Chinese Academy of China, Beijing, China*
*†These authors contributed equally to this work*
*\*Corresponding emails: hui.hui@ia.ac.cn and puxiang.lai@polyu.edu.hk*

**Abstract:**

Optical focusing and imaging through or inside scattering media, such multimode fiber and biological tissues, has significant impact in biomedicine yet considered challenging due to strong scattering nature of light. In the past decade, promising progress has been made in the field, largely benefiting from the invention of iterative optical wavefront shaping, with which deep-tissue high-resolution optical focusing and hence imaging becomes possible. Most of reported iterative algorithms can overcome small perturbations on the noise level but fail to effectively adapt beyond the noise level, e.g. sudden strong perturbations. Re-optimizations are usually needed for significant decorrelation to the medium since these algorithms heavily rely on the optimization performance in the previous iterations. Such ineffectiveness is probably due to the absence of a metric that can gauge the deviation of the instant wavefront from the optimum compensation based on the concurrently measured optical focusing. In this study, a square rule of binary-amplitude modulation, directly relating the measured focusing performance with the error in the optimized wavefront, is theoretically proved and experimentally validated. With this simple rule, it is feasible to quantify how many pixels on the spatial light modulator incorrectly modulate the wavefront for the instant status of the medium or the whole system. As an example of application, we propose a novel algorithm, dynamic mutation algorithm (DMA), which has high adaptability against perturbations by probing how far the optimization has gone toward the theoretically optimal performance. The diminished focus of scattered light can be effectively recovered when perturbations to the medium cause significant drop of the focusing performance, which no existing algorithms can achieve due to their inherent strong dependence on previous optimizations. With further improvement, the square rule and the new algorithm may boost or inspire many applications, such as high-resolution optical imaging and stimulation, in instable or dynamic scattering environments.

## 1. Introduction

Optical focusing through or within scattering media has been long desired yet considered challenging until the invention of feedback-based wavefront shaping [1]. In this technology, photons are phase-modulated by a spatial light modulator (SLM) before they are multiply scattered and become diffusive in the medium. A series of modulation patterns are displayed on the SLM under the control of an iterative optimization algorithm, and an optimal phase pattern is obtained when the feedback signal is maximized, indicating that the scattering-induced phase distortions are compensated as much as possible [2-4]. This technology has broad applications in biomedicine, for instance, deep tissue imaging [5-7], phototherapy [8], laser surgery, and photoacoustic imaging [9]. Moreover, wavefront shaping has been used to control the light transmission through a multimode fiber (MMF), in which speckles arise due to the intermodal interference and mode dispersion [10, 11]. MMF-based biomedical applications, such as endoscopic imaging [12, 13] and optogenetics [14, 15], are therefore progressively developed. Demonstrations based on MMF for wavefront shaping are therefore of interest since all these applications might be affected by the perturbations to the MMF on different levels, including consistently environmental noises (e.g. temperature change, pressure) and sudden strong perturbations (e.g. mechanical disturbances or biological motions). Those perturbations are due to the regular and/or irregular motion of the scattering medium or the system. Such instability is also a major factor that impedes wavefront shaping-assisted MMF from wider applications, since the modulated optical field through the MMF will accordingly decorrelate. The decorrelation essentially indicates that the transmission matrix (TM) of the MMF is highly susceptible to perturbations, such as bent, twisting, or temperature change [16, 17].

To combat against the instability, a careful designed setup can be utilized to separate the target photon from the perturbations due to the instable medium [18]. On the other hand, an efficient optimization algorithm for wavefront shaping is also required. Various algorithms have been reported and discussed, such as continuous sequential algorithm (CSA) [19, 20], stepwise sequential algorithm (SSA) [19], partitioning algorithm (PA) [19], genetic algorithm (GA) [21-24], particle swarm optimization (PSO) [25-27], simulated annealing algorithm (SA) [28, 29], and some artificial intelligence-assisted algorithms [30-32]. These methods have realized superior light focusing and even noise-resistance ability. But their optimization mechanisms originate from the numerical optimization without specific consideration about the physics behind the strong scattering process. These methods heavily rely on (or learns from) the net performance accumulated from previous iterations, especially for the GA, which needs a large pool containing many random phase/amplitude masks to initiate the optimization. The 'learning experience' is highly specific to the current status of the medium or the system, and hence the optimization can merely adapt and generalize to subtle instability of the medium, e.g. on the environmental noise level [21]. Once the perturbations to the medium are further strengthened beyond the noise level, e.g. a sudden perturbation, the transmission matrix of the medium may be altered significantly, and the resultant optical focus probably fades or even disappears. In this regard, another optimization is inevitable since the modulation patterns optimized in previous steps (without perturbations) are now weakly correlated with the new optimization condition that matches the state of the perturbated medium. An adaptive optimization algorithm is therefore desired, aiming to avoid strong dependency on the optimization from previous iterations.

Probably, the fact that existing algorithms are less adaptive to perturbations is the absence of a practical metric that can directly relate the instant focusing status with the accuracy of optimized wavefront. Such possibility is theoretically explored in this study based on the fully developed speckle patterns, which can be easily observed within or behind strong scattering media, such as an MMF or biological tissue. The intensity of a fully developed speckle pattern is governed by the negative exponential decay. As a result, phasors or elements in the corresponding TM of the medium follow a circular Gaussian distribution as discussed by Goodman [33]. Based on this plain assumption, we derive and define a metric called error rate (denoted as $r$), specific for the binary-amplitude modulation, to estimate how many pixels on the SLM are wrongly set based on the concurrently measured optical focusing: $r$ is physically related to the focusing performance as measured by peak-to-background ratio (PNR) by a simple square rule. This metric can imply how far the optimization has gone towards the theoretically optimal phase compensation, namely the ideal single-point focusing. Therefore, based on the real-time probed error rate instead of parameters inherited or accumulated from previous iterations, a novel algorithm called dynamic mutation algorithm (DMA) is developed in this study, as an application of the proposed practical metric. The optimization based on the error rate can automatically adapt strong perturbations: compared with other algorithms, the proposed DMA is advantageous, in both simulations and experiments, by its high adaptability and unique recovery ability against dynamic changes. Also, the diminished focus under perturbations (for example, twisting an MMF) can be effectively regained without additional operations, such as repeating the iterative optimization. With such guiding metric, the new algorithm may inspire further optimization of optical focusing in dynamic media from a practical perspective and boost more applications of wavefront shaping in living biological tissue.

## 2. Principle

The dynamic mutation algorithm (DMA) is an optimization method based on real-time experimental data instead of results from previous iterations, leading to high adaptability to dynamic changes. The key of the algorithm is to estimate the error rate of the corresponding amplitude mask, which implies how different the measured results deviate from the theoretical one, and then guide the optimization towards the optimal solution. In this section, we will elucidate the concept of the error rate and the DMA, followed by how the error rate can be used to modify the amplitude masks to achieve adaptive focusing. The steps involved in the optimization will also be explained.

### 2.1 The error rate and square rule

Beginning from the ideal focusing, the optimized wavefront, modulated by a binary-amplitude mask, is unique due to the deterministic transmission matrix of a medium ($T$), with elements $t_{mn}$. For simplicity, the input optical field is considered as a plane wave (with all phases set to zero) and modulated by a binary-amplitude only SLM, i.e. digital micromirror devices (DMD). Denoting the optimal wavefront $E_{op,1} = [e_1 \ \cdots \ e_N]^T$ as the optimal modulation for optical focusing at the 1$^{st}$ output channel, the optical field ($U_{f,1}$) with 1$^{st}$ output channel focused is governed by

$$U_{f,1} = TE_{op,1} = \begin{bmatrix} t_{11} & & t_{1N} \\ \vdots & \ddots & \vdots \\ t_{m1} & \cdots & t_{mn} & \cdots & t_{mN} \\ \vdots & & \ddots & \vdots \\ t_{M1} & & t_{MN} \end{bmatrix} \begin{bmatrix} e_1 \\ \vdots \\ e_n \\ \vdots \\ e_N \end{bmatrix} = \begin{bmatrix} u_{peak} \\ \vdots \\ u_{bg,m} \\ \vdots \\ u_{bg,M} \end{bmatrix} \quad (1)$$

where $u_{peak}$ and $u_{bg}$ represent the optical field of the focus (with peak intensity) and the non-focal regions (the background). Then, by dividing the $N$ ($N \gg 1$) input channels into two parts with ratio $r$ ($0 \leq r \leq 1$), i.e. $rN$ and $(1-r)N$, the peak intensity at the optical focusing ($u_{peak}$) can be formulated as

$$\left| u_{peak} \right|^2 = \left| \sum_{i=1}^{N} t_{1i} e_i \right|^2 = \left| \sum_{p=1}^{rN} t_{1p} e_p + \sum_{q=rN}^{N} t_{1q} e_q \right|^2 , \quad (2)$$
$$= \left| rN \langle t_{1p} e_p \rangle + (1-r)N \langle t_{1q} e_q \rangle \right|^2$$

where $\langle \rangle$ is the operator of ensembled average. Since the total input channels are randomly divided into two parts, the index $p$ and $q$ indicate the re-ordered elements for two divisions: $rN$ input channels, picked from the total $N$ input channels, are reordered with index $p$; the rest $(1-r)N$ input channels are reordered with index $q$. Considering the binary-amplitude modulation, the optimal amplitude of the elements in $E_{op,1}$ is determined by the first row in $T$ and element ($e_i$) and only turned 'on' if the real part of $t_{1i}$, denoted as $R_i$, is greater than zero:

$$e_i = \begin{cases} 1, & R_i > 0 \\ 0, & others \end{cases} \quad (3)$$

In this regard, both $\langle t_{1p} e_p \rangle$ and $\langle t_{1q} e_q \rangle$ positively contribute to the focusing with optimal modulation. Elements in $T$ are governed by the circular Gaussian distribution [33], i.e. real part ($R$) and imaginary part ($I$) of $T$ are statistically independent and follow the probability distribution density $f(x) = \exp(-x^2/2\sigma^2)/\sqrt{2\pi\sigma^2}$, where $\sigma$ is the standard deviation of the distribution. With Eq. (3), only elements in $T$ with positive real part ($R>0$) will be selected in the calculation. Since the imaginary part is independent against the real part, the selected elements with ($R>0$) have the imaginary part ($I$) fulfilling $-\infty<I<+\infty$. The terms of $\langle t_{1p} e_p \rangle$ and $\langle t_{1q} e_q \rangle$ from Eq. (2) can thus be expressed as

$$\begin{aligned}
N\langle t_{1p}e_p \rangle &= N\langle t_{1q}e_q \rangle = N\langle R \rangle + Nj\langle I \rangle \\
&= N\int_0^{+\infty} Rf(R)dR + Nj\int_{-\infty}^{+\infty} If(I)dI \\
&= N\int_0^{+\infty} \frac{R}{\sqrt{2\pi\sigma^2}} e^{-\frac{R^2}{2\sigma^2}} dR + Nj\int_{-\infty}^{+\infty} \frac{I}{\sqrt{2\pi\sigma^2}} e^{-\frac{I^2}{2\sigma^2}} dI \\
&= N \times \frac{\sigma}{\sqrt{2\pi}} + Nj \times 0 = \frac{N\sigma}{\sqrt{2\pi}}
\end{aligned} \quad (4)$$

where $j = \sqrt{-1}$. By substituting Eq.(4) into Eq.(2), the optimal peak intensity can be reduced as

$$I_{peak} = |u_{peak}|^2 = \left| r\frac{N\sigma}{\sqrt{2\pi}} + (1-r)\frac{N\sigma}{\sqrt{2\pi}} \right|^2 = \frac{N^2\sigma^2}{2\pi}. \quad (5)$$

Eq. (5) is consistent with previously reported studies [34, 35] with binary-amplitude modulation. Nevertheless, if the portion of $rN$ input channels are oppositely displayed, these input channels, with $R_i < 0$, are 'on' and negatively contribute:

$$\langle t_{1p}e_p \rangle = \langle R \rangle + j\langle I \rangle = \int_{-\infty}^{0} Rf(R)dR + j\int_{-\infty}^{+\infty} If(I)dI = -\frac{\sigma}{\sqrt{2\pi}} \quad (6)$$

Combining Eqs.(2), (5) and (6), the peak intensity with $rN$ incorrect-modulated input channel ($I_{peak,r}$) is revised as:

$$\begin{aligned}
I_{peak} = |u_{peak,r}|^2 &= \left| -r\frac{N\sigma}{\sqrt{2\pi}} + (1-r)\frac{N\sigma}{\sqrt{2\pi}} \right|^2 \\
&= (1-2r)^2 \frac{N^2\sigma^2}{2\pi}
\end{aligned} \quad (7)$$

In addition, the elements in TM are statistically independent [34] and therefore the optimal modulation for the 1st output channel (even with $rN$ incorrect input channel) does not affect the statistics of the other channels. Following the central limited law, the variance of $u_{bg}$ is $N/2$ times (the number of 'on' input channels) the variance of $t_{mi}$, ($m \neq 1$), i.e. $Var(t_{mi})$, so that the background intensity ($I_{bg}$) is expressed as:

$$I_{bg} = \langle |u_{bg}|^2 \rangle = Var(u_{bg}) = \frac{N}{2} Var(t_{mi}) = \frac{N}{2} 2\sigma^2 = N\sigma^2 \tag{8}$$

Finally, a relative peak to background ratio (PBR), denoted as $\eta'$, due to the r-ratio incorrected modulation can be obtained by

$$\eta' = \frac{\eta_r}{\eta_0} = \frac{I_{peak,r}/I_{bg}}{I_{peak}/I_{bg}} = \frac{(1-2r)^2 N/2\pi}{N/2\pi} = (1-2r)^2, \tag{9}$$

where $\eta_0$ ($\eta_0 = N/2\pi$) is the theoretical PBR and $\eta_r$ ($\eta_r = (1-r)^2 N/2\pi$) is the PBR with r-ratio of pixels incorrectly modulating the input wavefront. For simplicity, the relationship indicated by Eq. (9) is termed as the "square rule" for binary-amplitude modulation. Practically, the square rule can be generalized to the case that the r-ratio of pixels oppositely modulate the wavefront in any given modulating masks rather than only the optimal mask. This generalization will be proved experimentally in Section 2.5.

Mathematically, the relative PBR is essentially related to the Pearson's correlation coefficient ($\rho_r$) between the optimal mask ($E_{op,1}$) and a mask with r-ratio incorrect modulation ($E_{r,1}$). Their elements, i.e. $e_i$ and $e_{r,i}$ for $E_{op,1}$ and $E_{r,1}$ respectively, obey to the symmetric Bernoulli distribution, i.e. e ~ Bernoulli (p=0.5) so that both $\langle e_i \rangle$ and $\langle e_{r,i} \rangle$ are 0.5 and $\langle e_i e_{r,i} \rangle = 0.5(1-r)$. By defining $\delta f = f - \langle f \rangle$, the $\rho_r$ can be formulated as

$$\rho_r = \frac{\langle \delta e_i \times \delta e_{r,i} \rangle}{\sqrt{\langle \delta e_i \rangle^2 \langle \delta e_{r,i} \rangle^2}} = 4 \langle e_i e_{r,i} \rangle - 1 = 1 - 2r. \tag{10}$$

With Eqs. (9) and (10), the ratio ($r$) of the opposite modulating channels, or termed as the error rate, can be directly estimated from the experimental PBR ($\eta_r$) via a simple relationship if the range of error rate is limited below 0.5:

$$r = (1 - \sqrt{\eta'})/2 \tag{11}$$

Notably, the value of $r$ can increase above 0.5, and the $\eta'$ will increase accordingly as indicated by Eq. (9), which is symmetric regarding to $r = 0.5$. For $r > 0.5$, the focusing performance will be equivalently reverses the selection of Eq. (3): the element ($e_i$) in the optimal mask is turned 'on' for $R_i < 0$, and the effective error rate becomes (1-r). It is because turning 'on' the elements for either $R_i < 0$ or $R_i > 0$ shares equivalent performance for optical focusing according to the symmetry of the circular Gaussian distribution. Therefore, considering Eq. (3), it is straightforward to use the Eq. (11) ($r \leq 0.5$) to estimate the ratio of incorrectly modulating pixels.

Experimentally, the estimated error rate ($r$) of any optimized pattern is obtained as following: an $N$-element DMD is used to modulate the input wavefront and then a 'theoretical PBR' is calculated by $\eta_0 = N/2\pi$; the 'experimental PBR' ($\eta_{ex}$) is obtained from the instant speckle pattern, i.e. a focus at the target position; finally, the experimental error rate for each focusing optimization can be obtained by substituting Eq.(10) into the Eq. (11):

$$r = (1 - \sqrt{\eta_{ex}/\eta_0})/2 = (1 - \sqrt{2\pi\eta_{ex}/N})/2 \tag{12}$$

**2.2 Mutation rate**

Benefit from Eq. (9), the percentage of DMD elements with incorrect modulation in the mask can be directly estimated. And an ideal solution, or mask, can be obtained if the incorrect elements are corrected. That said, the exact positions/indexes of these elements are unknown. Inspired by the mutation process in GA, having a suitable mutation rate to change the state of modulating elements regarding the error rate may be able to improve the optimization.

In the GA, the mutation rate is usually preset in a decaying manner and it gradually becomes smaller regardless of the actual optimization performance or status. In the proposed DMA, the mutation rate is adjusted dynamically according to the error rate so that the information about how instable medium instantly affects the optimized mask can be considered. And one more benefit of integrating the error rate is that: in every iteration, the number of elements to be mutated ($N\mu$) can be well controlled below the number of the wrongly modulating elements ($Nr$), i.e. $N\mu \lesssim Nr$. As an example to scale the mutation rate, the mutation rate ($\mu^{(s)}$) in the $s^{th}$ iteration can be simply set to be proportional to the instant error rate ($r^{(s)}$) with a mutation constant ($C$) that is greater than unity (Eq. 13). To generate new DMD patterns in the $s^{th}$ iteration, a total number of $N\mu^{(s)}$ elements in the DMD mask generated in the $(s-1)^{th}$ iteration are randomly selected and mutated by reversing the element state of on/off or equivalently following Eq.(14).

$$\mu^{(s)} = r^{(s)}/C \tag{13}$$

$$e^{(s)} \leftarrow 1 - e^{(s-1)} \tag{14}$$

The function with respect to the mutation constant, Eq.(13), is to bound the mutation rate between 0.5/$C$ and 0, if the error rates at the beginning and at the end of the optimization are assumed to be 0.5 and 0, respectively. The mutation constant is the only parameter needed to be set before the optimization; a smaller mutation constant is suggested in unstable environments to provide a larger range of mutation rates. The mutation rate is autotuned by the algorithm within the range in response to the actual situations, so as to increase the chance for the incorrect elements to be mutated and lead the optimization towards the theoretical result.

## 2.3 The DMA workflow

The block diagram in Fig. 1 shows the typical workflow of the DMA for the binary-amplitude modulation with DMD. First, all DMD pixels are set to be one ("on" state). The error rate is found according to Eq. (12) and the mutation rate is computed through Eq. (13). Then the mask is mutated to generate a new DMD pattern with Eq. (14). Then, the error rate is assessed again based on the instantly measured focus performance, i.e. the PBR If the error rate becomes smaller, which means the PBR is improved, a new mutation rate is calculated according to the error rate. If there is an abrupt rise of the error rate, probably caused by the changes of medium of interest, the mutation rate will be updated as well. Otherwise, if the error rate is just slightly fluctuating and shows no improvement in PBR, the current mask is mutated again with the same mutation rate. The mutation rate is not updated in this case in order to minimize the error rate fluctuations. This process is repeated until the PBR of optical focusing saturates or plateaus.

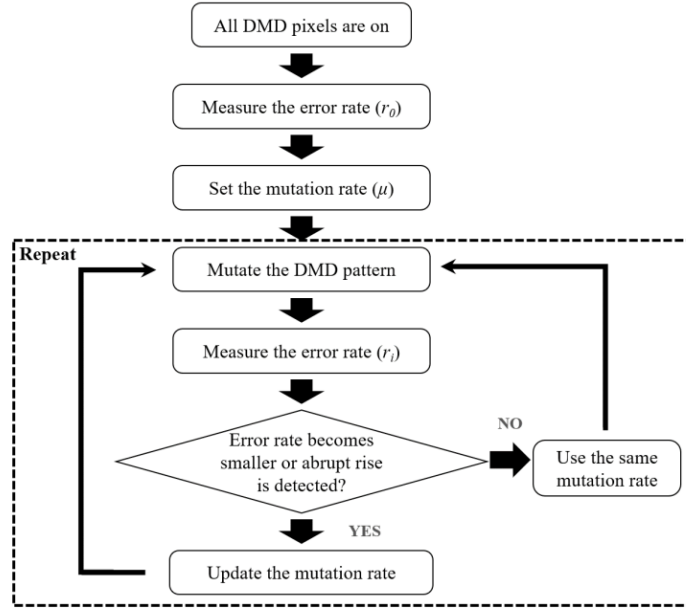

Fig. 1 Block diagram showing the workflow of the dynamic mutation algorithm (DMA)

## 2.4 Experimental setup

The experimental setup used in this study is shown in Fig. 2a. A continuous wave 532 nm laser source (EXLSR-532-300-CDRH, Spectra Physics, USA) is used to illuminate the DMD (DLP4100, Texas Instruments Inc., USA). A pair of convex lenses ($L_1$ and $L_2$) are used to expand the light beam, such that it covers all pixels on the DMD. Another pair of convex lenses ($L_3$ and $L_4$) are used to de-magnify the beam after it is modulated by the DMD. After that, the shrunk modulated light is focused with a 40x objective lens (NA=0.65) onto a scattering sample. A CMOS camera (Blackfly S BFS-U3-04S2M-CS, FLIR, Canada) is placed behind the scattering sample and an image is captured in each measurement for the calculation of the instant PBR and the error rate. The PBR can be calculated by dividing the intensity of the target mode by the average background intensity. A 1-meter bare optical multimode fiber (MMF) (SUH200, Xinrui, China, with diameter =200 μm, NA=0.22) is chosen as the scattering sample, with two collimators (PAF2-A4A, Thorlabs, USA) and a fiber rotator (HFR007, Thorlabs, USA). During the optimization for optical focusing, 64×64 input modes are used (16×25 pixels on the DMD are grouped as a mode and each pixel is 10.8 μm × 10.8 μm), and every algorithm is run for 10,000 measurements without stop even rotation to the fiber is applied.

In the following sections, the parameters of the investigated algorithms used for the simulation and experimental are set be the same as follows: for the DMA, the mutation constant in Eq. (13) is set to 200. For the GA, the population size is 20 and the offspring size is 10. The initial mutation rate is 0.1, which decays exponentially to a final value at 0.001. For the CSA, there is no preset parameter, whose input modes are optimized one by one with a linear rastering manner [19, 20]. These initial parameters related to specific algorithms are summarized in the Table 1. All these algorithms are implemented for 10000 measurement and each measurement is set to spend 0.2 second.

As an example, speckle before and after DMA optimization are showed in Fig. 2. Fig. 2b shows the speckle field before optimization, where the PBR at the target position (central point) is around 3. After wavefront shaping optimization guided by the DMA, the focus has a PBR enhanced to 120 as shown in Fig. 2c. The full-width half-maximum (FWHM) of the focus is 15.2 and 14.5 μm in the horizontal and vertical directions, respectively.

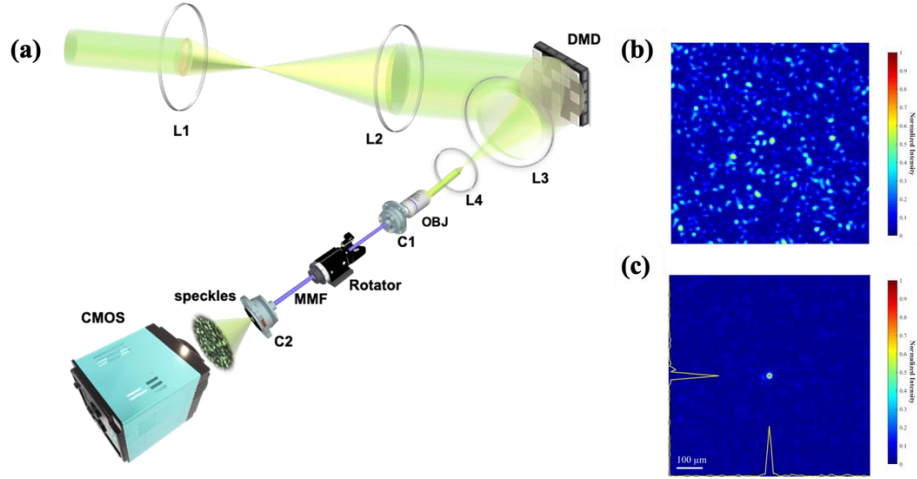

Fig. 2 a) Experimental setup. $L_1$: f=60 mm; $L_2$ & $L_3$: f=250 mm; $L_4$: f=50 mm; DMD: 1920x1080 digital micromirror device; OBJ: 40x objective lens (NA=0.65); MMF: 1-meter multimode bare optical fiber (diameter =200 μm, NA=0.22). A fiber rotator was added in the fiber rotation experiment. $C_{1-2}$ are the fiber collimators. b) Speckle field before optimization. c) Focus formed after optimization by the DMA (PBR=120). The yellow lines indicate the profiles of the focus along the horizontal and vertical directions, respectively.

Table 1 Initial parameters set for DMA, GA, and CSA

| Algorithm | DMA | GA | CSA |
|---|---|---|---|
| **Initial parameter** | Mutation constant =200 | 1. Population size = 20<br>2. Offspring size = 10<br>3. Initial mutation rate = 0.1<br>4. Final mutation rate =0.001<br>5. Decay constant = 200 | NONE |

## 2.5 Verification of the square rule

Numerical and experimental proof of the square rule have been done to validate its practical implications. The square rule can be simple and straightforward (denoted as theoretical prediction in Fig. 3a-b). And it can be computationally recreated in simulation by using a TM whose elements follow the circular Gaussian distribution (denoted as ideal simulation in Fig. 3b). Yet, to prove the square rule experimentally, a series of error rates are selected, i.e. 0%, 10%, 20%, 30%, 40% and 50%, and each data point is the average of five executions. First, we use a TM-based method [36] to generate an optimal DMD mask (set as r=0) for an optimized focus through a MMF and then mutate the optimal mask with r-ratio of pixels to test the corresponding focusing performance, i.e. PBR. As shown in Fig. 3a, the experimental η′-r curve is shaped like a parabola with a right shift away from the theoretical curve, $η′= (1−2r)^2$. Such shift may be attributed to the instability of the MMF, since instability effect can be accumulated during the process of TM measurement. The measured TM may be subjected to deviation from the ideal assumption that the elements in TM of the scattering sample follows an ideal circular Gaussian distribution. In view of this, the real part of the measured transmission matrix of the MMF is found to have a right shifted Gaussian distribution (mean=0.002 and standard deviation=0.04) (inset in Fig. 3a). Based on this TM measurement, a TM, whose real part follows the Gaussian distribution with mean=0.002 and standard deviation=0.04, is generated to form an optimized focus with a series of error rate to degrade the performance. The induced η′-r curve (denoted as simulation in Fig. 3a) matches well with the experimental one. Notably, in our experimental setup, the TM measurement needs around 30 min to complete, and the obtained real part of TM is used to generate the optimal mask. Decorrelation of the medium is hardly ignored and inevitably coupled into the measured TM. As well, all the mutated masks (with r=10~50%) are generated from the same optimal mask after the TM measurement is completed. These masks therefore carry the information of medium instability from the TM measurement. Furthermore, these masks are sequentially displayed on the DMD to modulate the wavefront, which also costs time. When the case of r=50% is tested, the medium has been altered and decorrelated from the status when the case of r=0 is tested. This may imply that the right shift of the experimental η′-r curve, or the unsatisfactory optimization result, can be attributed to the instable medium represented by a biased or shifted TM.

Nevertheless, the instable medium (on the noise level) can still be governed by the square rule to some extent, but the accuracy fails to maintain. That is because the square rule is based on an unchanged and stable medium. Therefore, to experimentally recreate the square rule, at least, the time span between the generation of the optimal mask and mutated mask is limited within the decorrelation window of the medium. For example, an optimal mask is generated for every error rate investigation during experiment. By doing so, the instable effect due to the practical medium can be almost eliminated, as shown in Fig. 3b, where the η′-r curve attained from experiments matches well with the theoretical curve as well as the ideal simulation curve. Therefore, the square rule functions well in a real-time representation, and in other words, the error rate can provide an effective instant metric to evaluate the 'distance to the ideal optimal optimization'. That will provide a plain yet universal perspective to analyze the imperfect focusing performance for the scattering medium, even it is heavily perturbated.

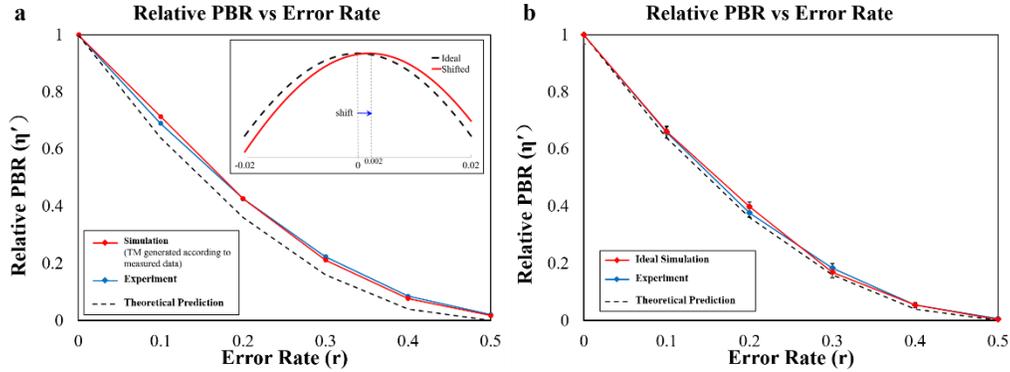

Fig. 3 Relative PBR-error rate curves (η′-r curve). The curve based on the theoretical prediction following the derived square rule is plotted in dashed line in both a) and b). In a): the blue-diamond line shows the optimization based on the TM measured at *r*=0, and the instable effect is included without re-measuring the TM for other error rates; the red-diamond line is based on the simulation with a TM, whose real part distribution is subjected to a shift (the mean of the Gaussian distribution is shifted from 0 to 0.002 in the inset). In b): the optimal mask for each investigated error rate is re-obtained to eliminate the instable effect in experiments (blue-diamond line) and simulations (red-diamond line), matching well with the theoretical η′= $(1−2r)^2$ curve. Relative PBR for each error rate was repeated for 5 executions and the error bars show the standard deviation of the measurements.

## 3. Results

### 3.1 Simulation

Simulations are done to evaluate the performance of the DMA, which is compared with two representative existing algorithms, i.e., the genetic algorithm (GA) and continuous sequential algorithm (CSA). In addition to their popularity, the GA and CSA are selected as they share some similarities with the DMA. Both the DMA and GA have a mutation process and target at optimization in unstable environments. Meanwhile, the DMA and CSA are straight-forward algorithms that do not rely much on previous results. Simulations with the DMA, GA and CSA have been performed under various conditions of different levels of noise, and the results are compared based on the PBR throughout a fixed number of measurements (or iterations). Whenever the intensity of the target mode is measured, it is counted as one measurement and GA usually needs several measurements for one iteration. Each curve in the plots is an averaged result of 50 executions with a new transmission matrix generated to simulate the scattering process for every execution. $N= 64\times64$ input modes (modulating elements for binary-amplitude modulation) are used and the output mode at the center is chosen for optimization.

### 3.1.1 Influence of noise level

In this section, the algorithms are compared under different levels of noise. Additive white Gaussian noise is added in every intensity measurement to mimic the instability of the optical system in actual environment [21]. The Gaussian noises, with standard deviations of 30% and 60% of the initial average intensity $\langle I_0 \rangle$, are set to represent the low-noise and high-noise situations, respectively.

Fig. 4a-c shows the simulation results under the conditions of noise-free, low-noise ($0.3\langle I_0\rangle$) and high-noise ($0.6\langle I_0\rangle$), respectively. At noise-free and low-noise situations, the DMA has the fastest growth of the initial PBR and achieves high PBR. Although the CSA obtains the highest final PBR in the noise-free case, its performance declines drastically with increased noise level. At high-noise situation, the GA, benefitting from its large population of optimizing masks, exhibits its superior noise resisting ability. Meanwhile, the DMA can also reach a comparable level of PBR without such a large population presented in GA. Wavefront optimization guided by the error rate therefore is immune to the need of a large population.

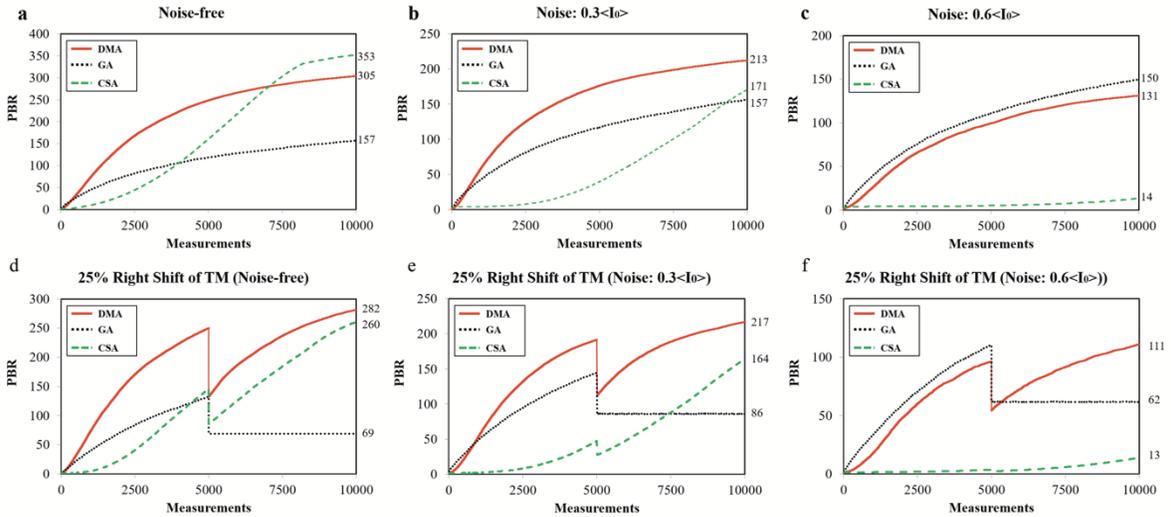

Fig. 4 Simulation results of the DMA, GA and CSA under different conditions. a) Noise-free; b) low-noise: 0.3 <$I_0$>; c) high-noise: 0.6 <$I_0$>; c)-f) 25% right shift of the transmission matrix (at the 5000th measurement) applied to noise-free, low-noise and high-noise conditions.

### 3.1.2 Influence of transmission matrix change

Apart from the noise caused by the instability of the optical system, optimization results can be greatly affected by the instability or slight movement of the scattering medium. To simulate this situation, a 25% right shift of the scattering medium was implemented at the 5000[th] measurement. The scattering medium was represented by a *M* output modes × *N* input modes transmission matrix. *M* × 0.25*N* new elements, following the same circular Gaussian distribution, are generated and inserted to the left of the matrix. The right 25% of the original matrix elements are removed so that a new *M*×*N* transmission matrix to mimic the shift of the medium is formed. Different from the noise addition process, which only affects the intensity measurement, the shift also leads to changes in the transmission matrix. Simulations are done in noise-free, low-noise and high-noise situation. The simulation parameters used for the algorithms here are the same as those mentioned in Section 3.1.1. Fig. 4d-f shows how the DMA, GA, and CSA respond to the shift in transmission matrix.

As seen, right after the transmission matrix is changed, there is a sudden drop of PBR in all three algorithms. The GA fails to adapt the sudden change of the TM shift for all the three investigated noise conditions, which is associated with the mechanism of the GA: the offspring (amplitude masks) with lower cost (intensity) in the population is replaced and the best offspring is always kept [21]. Also, the whole population is generated based on the medium status before the TM shift, whose dependency and the correlation regarding the largely shifted TM is relatively low. Therefore, when it encounters a relatively large enhancement drop, it is hard to produce offsprings with cost (i.e. the PBR) larger than the former best one, which makes the optimization trapped in a local maximum. In constrast, DMA and CSA successfully recover the focus after the TM shift under noise-free and low-noise conditions. Such achievement is probably due to their absence of a pool with a large population, whose information is strongly related to the status of the medium before the TM shift. Or equivalently, the population size of the DMA and CSA is one so that the modulating mask can be instantly guided by the information from the sudden shift without constraints from the other masks in the pool. Under the high-noise condition, the DMA is the only algorithm that can adapt to the sudden change and bounce back to the original level after ~5000 measurements. For CSA, limited by its weak noise resisting ability, fails to tackle the high-noise conditions.

As comparison, the square rule, providing the DMD error rate from the instant PBR, shows its advances to deal with different levels of noise conditions and sudden changes. In the next section, experimental performance will be further discussed.

### 3.2 Experiment

*3.2.1 Focusing against strong noise*

With external perturbations on the noise level, the optimizations for single point focusing via the DMA, GA, and CSA are shown in Fig. 5.The final PBR the algorithms achieved can be similarly divided into two groups, i.e. DMA and GA with effective focusing and CSA with weak effectiveness, compared to the simulation results at situation with noise of $0.6\langle I_0\rangle$ (Fig. 4c). Both the DMA and GA demonstrate their robustness in noisy environment. The DMA shows a higher initial PBR rate and reaches its optimal state after around 5000 measurements. The GA transcends the DMA at around 7500[th] measurement. The CSA has a slow initial PBR rate as the optimization starts from the pixels at the edge of the mask, which contributes less to the optimization due to the Gaussian beam used in experiment. The contribution from the modulating element increases when it comes to the central part of the mask, then slows down again and eventually reaches its maximum when the process approaches another end of the mask. The CSA is sensitive to strong noise [27], so it cannot obtain an PBR as high as the other two algorithms.

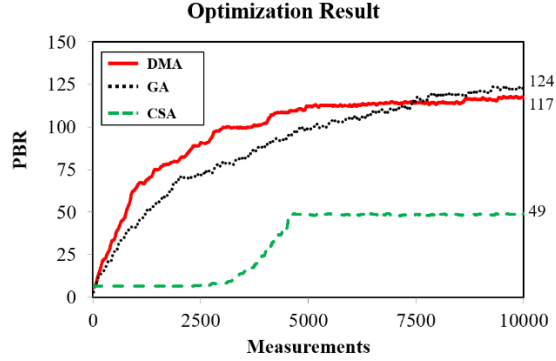

Fig. 5 Experimental results of the DMA (red solid curve), GA (black dashed curve), and CSA (green dotted curve) focusing performance against strong noise.

*3.2.2 Focusing against strong perturbations*

With more apparent perturbations, e.g. a slight movement, a bending, or a small rotation, to the MMF, the corresponding TM can be significantly changed and the speckle patterns decorrelated. If that occurs during the experiment, the optimization process is disrupted, and the resultant focal spot may be ruined. In this section, experiments were done to study how perturbations, with rotation to the MMF as an example, affect the optimization and how different algorithms respond to such heavy instability. The same experimental setup was used with an additional fiber rotator, which can rotate the MMF with various angles.

The relationship between the degree of rotation to the fiber and the corresponding PBR drop is shown in Fig. 6a. Different degrees of rotation (2.5°, 5°,7.5°, 10°, 12.5° and 15°) were introduced when the PBR reached 100. And notably, the optimization algorithms are kept running during the whole optimization and not commanded to stop before and/or after the rotations. As the MMF is altered by the rotation, the PBR drops immediately, as seen in Fig. 6c. Moreover, the more the fiber is rotated, the larger the percentage drop for the PBR. It indicates that the corresponding TM is altered significantly due to the fiber rotation. By applying DMA to optimize the focus, the optimization can automatically adapt the degraded focus without re-running the optimization process. As shown in Fig. 6c, the DMA adapts to the focus degradation with various degrees of fiber twisting, i.e. 2.5°, 5°and 7.5° rotation corresponding to ~20%, ~40% and ~60% PBR drop. DMA is always able to recover PBR to the value before perturbation regardless of how much the PBR has dropped. And Fig. 6b shows the number of measurements required for the PBR to rebound to the level before perturbations is increasing with respect to the rotation angle. It again validates that strong perturbations can significantly change the status of the medium, which poses challenges to any iterative algorithm. The DMA, on the other hand, shows critical advantages over current popular algorithms.

As an example to study how the DMA, GA and CSA combat against the heavy instability of the MMF, a 5° rotation for the MMF was implemented at the $5000^{th}$ measurement for these three algorithms. Fig. 7 shows how the algorithms perform throughout the experiments and Fig. 7 shows the focal spots before optimization ($0^{th}$ measurement), right before fiber rotation ($5000^{th}$ measurement), right after fiber rotation ($5001^{st}$ measurement), and after re-optimization ($15000^{th}$ measurement). The experimental results agree well with the simulation results in Section 3.1 that the DMA and CSA show their recovery abilities. The DMA rebounds soon after the rotation and takes around 4000 measurements to regain the PBR it has achieved before the rotation, and the resultant optical focus is as bright as, if not brighter than, the focus before perturbation (Fig. 8). Comparably, the CSA does not recover right away after the rotation. It starts to recover after around 6250 measurements.

The recovery efficiency of the CSA may depend on when the perturbation occurs. In experiment, as the perturbation is induced when the optimizing elements are near the edges of the DMD, the recovery speed is slow. More importantly, merely changing one element for each measurement in CSA is not efficient to overcome the instability since the positive contribution from one element is probably below the noise level. In contrast, the GA cannot obtain further PBR after the rotation as the optimization is trapped in the local maximum due to three reasons. First, the optimizing masks in the population library is merely based on the medium status before perturbations. Second, the decorrelation due to 5° rotation cannot be tolerated or generalized from that population generated via GA. Third, the mutation process in GA is not adaptive to the sudden perturbations during optimization since the mutation rate in GA is exponentially decayed regardless of the focus degrading. Collectively, the DMA can effectively battle those defects inherently embedded in the CSA and GA, which are also shared by most of the popular algorithms.

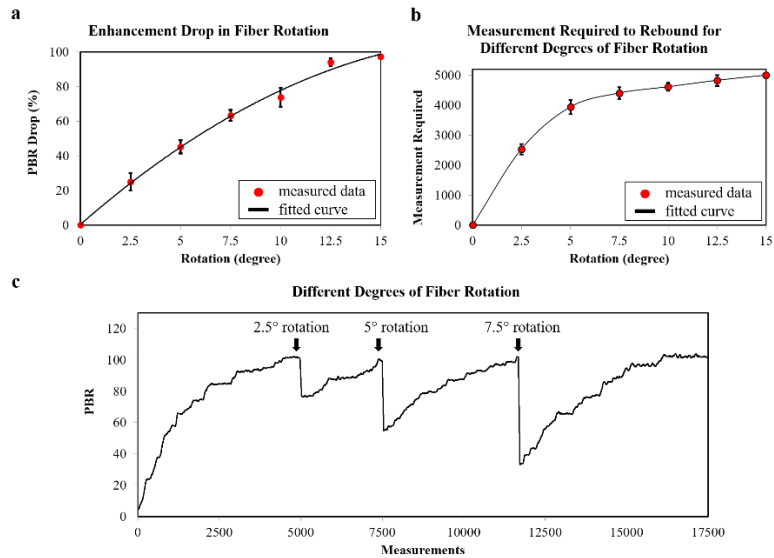

Fig. 6 a) Relationship between fiber rotation and PBR drop. b) Measurement required to rebound for different degrees of fiber rotation. Each dot in a) and b) is averaged from 5 executions and the error bars show the standard deviation of the measurements. Optimizations in a) and b) are realized by the DMA. c) Experimental focusing performance of the DMA in response to 2.5°, 5°, and 7.5° fiber rotation.

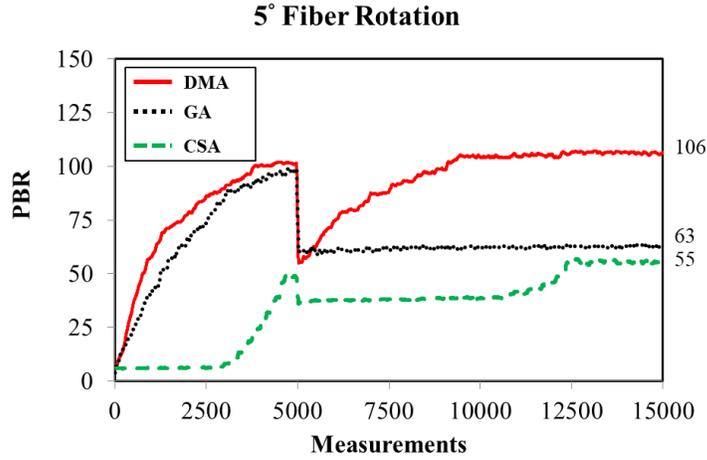

Fig. 7 Focusing performance of the DMA, GA, and CSA in response to 5° fiber rotation.

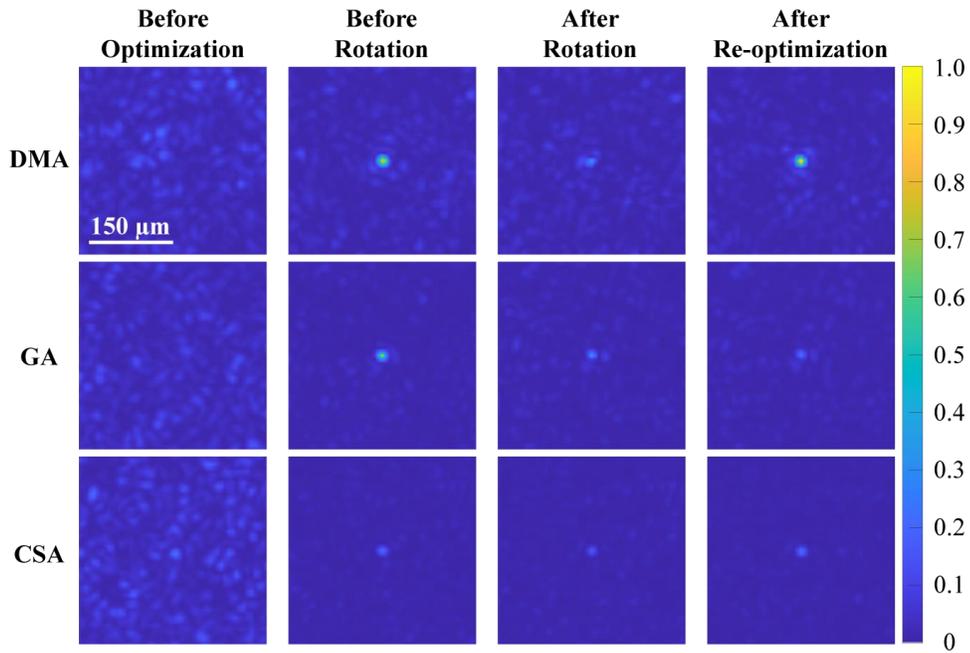

Fig. 8 Focal spots at 4 different stages with different algorithms: before optimization (0th measurement), before fiber rotation (5000th measurement), right after 5° fiber rotation (5001st measurement) and after re-optimization (15000th measurement). The 150 μm scale bar is applicable to all images in this figure.

## 4. Discussion

As seen, the simulation and experiment results have demonstrated the high adaptability of the DMA. It performs comparably with GA in noisy environment and overcomes the heavy perturbations with a robust recovery ability. Apart from this distinctive adaptability, the ease of implementation is another advantage of the DMA. For the GA, several key parameters, such as population size, offspring size, and mutation rates, are needed to be adjusted appropriately at the beginning [21]. However, in DMA, only one parameter is required to be set, which is the mutation constant, a number bounding the mutation rate. Benefit from the error rate and the square rule, the optimization error in the modulating mask can be easily quantified, causing the whole optimization process to be straightforward. It is just simply based on real-time measurements, and less likely to be affected by the improper selection of parameters. Also, the error-rate based DMA does not strongly depend on the modulating mask in previous iterations and, therefore, the adaptability can be effectively achieved to deal with dynamic media. As an example of the square rule's applications, the DMA does show its capability to adapt to strong and/or sudden perturbations benefitting from the use of the practical metric, the error rate. Note that such metric can also be incorporated into other optimization methods, such as GA, to improve the adaptability.

Although only one of the numerical optimization methods, i.e. GA, is chosen for demonstration in this study, others, such as SA and PSO, etc., are similar. Including their pools with a large population of masks, the mechanisms to generate new modulating patterns originate from the philosophy of numerical optimizations: 1) the portion of the modulating elements to be mutated is preset or decay with certain rules; 2) the monitored PBR are used to produce an acceptance probability of the new generated masks or to update the mask. These two mechanisms ensure the generalization to the noise, even on the scale of $<I_0>$ [27], around a specific equilibrium of the medium. However, a sudden strong perturbation behaves differently since the equilibrium of the medium has been considerably changed, experience based on the previous equilibrium is 'out-of-date' as discussed in last section. Therefore, considering the physics-based square rule probably enhance the adaptability of the existing methods. Notably, if a new parameter is incorporated in those methods, other parameters probably need to be further tuned to match the function of the square rule, which is a non-trivial manipulation. Incorporation of other optimization algorithm with the square rule is therefore beyond the manuscript.

## 5. Conclusion

In this study, a simple square rule of binary-amplitude modulation based wavefront shaping optical focusing based on universal strong scattering media has been theoretically obtained. With this rule, the real-time error in the modulating mask can be simply calculated from the concurrently measured PBR of the optical focus. Based on such a real-time metric, a novel feedback-based wavefront shaping algorithm, dynamic mutation algorithm (DMA), has been proposed. Both the simulation and experiment results have demonstrated its high adaptability and unique recovery ability that no other existing algorithms can achieve: focusing of diffused light can be regained without re-running the optimization even after a 60% drop of the PBR. It is due to the application of the square rule, which guides the optimization with the universal physics knowledge about the strong scattering process instead of a random guess. Notably, the square rule assumes that the transmission matrix of a medium follows a circular Gaussian distribution. It can be easily fulfilled when the transmitted medium is a strong scattering medium: photons are multiply-scattered and most of these scattering events are independent [33]. Therefore, the square rule between the DMD error rate and the degraded focus can be generally applied to process in strong scattering regime. The algorithm is therefore particularly suitable to be used in heavily unstable or motional scattering environments. Note that the DMA in this study merely serves as an example application to utilize the error rate and square rule to optimize the single-point focusing. On other hand, MMF is used as the example of scattering media in this study, so that by applying rotation of certain degrees we can induce controllable, repeatable, and quantifiable perturbations to the resultant speckle patterns. This is necessary for the current phase of proof of principle, although not ideal. With further improvement, we believe the study may boost or inspire many applications of wavefront shaping with instable media or even living biological tissue.

**Acknowledgement**

This work was supported in part by the National Key Research and Development Program of China (2017YFA0700401), National Natural Science Foundation of China (NSFC) (81671726, 81930048, 81627805, 81827808), Hong Kong Research Grant Council (25204416), Hong Kong Innovation and Technology Commission (ITS/022/18, GHP/043/19SZ, GHP/044/19GD), Guangdong Science and Technology Commission (2019A1515011374, 2019BT02X105), Shenzhen Science and Technology Innovation Commission (JCYJ20170818104421564), and the Youth Innovation Promotion Association of Chinese Academy of Sciences (2018167).

**Disclosures**

The authors declare no conflicts of interest.